\documentclass[apjl]{emulateapj}
\usepackage{graphicx,epsfig}
\usepackage{txfonts,amssymb}
\usepackage{color,xcolor}
\usepackage{natbib}
\DeclareMathAlphabet{\mathitbf}{OML}{cmm}{b}{it}
\DeclareMathAlphabet{\mathf}{OML}{cmm}{c}{sl}
\renewcommand{\vec}[1]{{\mathitbf #1}}
\newcommand{\Bvec}{{\vec B}}
\newcommand{\Bz}{{\mathf B_z}}
\newcommand{\Bh}{{\mathf B_h}}

\newcommand{\jvec}{{\vec j}}
\newcommand{\jz}{{\mathf j_z}}

\newcommand{\ie}{i.\,e.}
\newcommand{\eg}{e.\,g.}

\newcommand{\sdo}{{\it SDO}}
\newcommand{\stereo}{{\it STEREO}}
\newcommand{\rhessi}{{\it RHESSI}}
\newcommand{\goes}{{\it GOES}}
\newcommand{\white}{\color{white}}
\newcommand{\black}{\color{black}}

\slugcomment{Accepted for publication in \apj\ on May 11, 2016}
\begin{document}

\title{Temporal and spatial relationship of flare signatures and the force-free coronal magnetic field}
\author{J.~K.~Thalmann\altaffilmark{1}, A.~Veronig\altaffilmark{1,2}, Y.~Su\altaffilmark{1,3}}
\email{julia.thalmann@uni-graz.at}
\altaffiltext{1}{Institute of Physics/IGAM, University of Graz, Universit\"atsplatz 5/II, 8010 Graz, Austria}
\altaffiltext{2}{Kanzelh\"ohe Observatory for Solar and Environmental Research, University of Graz, Universit\"atsplatz 5/II, 8010 Graz, Austria}
\altaffiltext{3}{Key Laboratory of Dark Matter \& Space Astronomy, Purple Mountain Observatory, Chinese Academy of Sciences, 2 West Beijing Road, 210008 Nanjing, China}

\begin{abstract}
	We investigate the plasma and magnetic environment of active region NOAA~11261 on 2 August~2011 around a \goes\ M1.4~flare/CME (SOL2011-08-02T06:19). We compare coronal emission at (extreme) ultraviolet and X-ray wavelengths, using \sdo\ AIA and \rhessi\ images, in order to identify the relative timing and locations of reconnection-related sources. We trace flare ribbon signatures at ultraviolet wavelengths, in order to pin down the intersection of previously reconnected flaring loops at the lower solar atmosphere. These locations are used to calculate field lines from 3D nonlinear force-free magnetic field models, established on the basis of \sdo\ HMI photospheric vector magnetic field maps. With this procedure, we analyze the quasi-static time evolution of the coronal model magnetic field previously involved in magnetic reconnection. {This allows us, for the first time, to estimate the elevation speed of the current sheet's lower tip {\it during} an on-disk observed flare, as a few kilometers per second.} Comparison to post-flare loops observed later above the limb in \stereo\ EUVI images supports this velocity estimate. {Furthermore, we provide evidence for an implosion of parts of the flaring coronal model magnetic field, and identify the corresponding coronal sub-volumes associated to the loss of magnetic energy. Finally, we spatially relate the build up of magnetic energy in the 3D models to highly sheared fields, established due to dynamic relative motions of polarity patches within the active region.}
\end{abstract}


\section{Introduction}

Magnetically complex active regions (ARs), including those exhibiting a $\delta$-configuration, are known to tend to higher flare and CME productivity (see reviews by, \eg, \cite{lrsp-2008-1,2009SSRv..144..351V} and see also, \eg, \cite{2000ApJ...540..583S}). Following the general thoughts of the standard model of eruptive flares \citep{1964NASSP..50..451C,1996ASPC...95...42S,1974SoPh...34..323H,1976SoPh...50...85K}, a vertical current sheet is formed behind an outward moving CME which drags the bipolar magnetic field of the erupting coronal structure along \citep[for a review see][]{2002A&ARv..10..313P}. The resistivity in the current sheet may be locally enhanced so that magnetic field and plasma are decoupled, allowing the magnetic field configuration to change within a localized diffusion region. This is accompanied with the release of previously stored magnetic energy via the rapid dissipation of stored electric currents \citep[for recent reviews see][]{2011LRSP....8....6S,2015SoPh..290.3425J}.

An important feature of such events are apparent chains of enhanced low-atmosphere emission, preferentially observed in the chromosphere and transition region, so-called flare ribbons. They develop when non-thermal particles accelerated at the coronal reconnection site, somewhere along the current sheet, are trapped and guided along magnetic lines of force (field lines) towards the low atmosphere (\eg, \cite{2001SoPh..204...69F} and for a review see \cite{2011SSRv..159...19F}). Since the current sheet is a horizontally elongated region, particles may be simultaneously accelerated at multiple positions along of it. Consequently, the arrival of the accelerated particles along neighboring field lines in the low atmosphere may be nearly simultaneous, resulting in the observed thread-like enhanced emission.

Since flare ribbons mark the locations where newly reconnected field lines connect to the lower atmosphere, they are located in opposite-polarity regions on either side of a polarity inversion line (PIL). As reconnection occurs successively higher in the solar corona, the newly reconnected field lines close down farther away from the PIL, resulting in an apparent increasing separation of the flare-ribbon emission. Importantly, the magnetic field rooted in the outer edges of the flare ribbons maps back to the coronal reconnection site, more precisely, to the lower tip of the current sheet \citep[\eg,][]{2004SoPh..222..115L}. The plasma that is heated by the arriving flare-accelerated non-thermal particles at the low-atmosphere field lines' footpoints emits hard X-rays (HXRs). {Once heated, the low-atmosphere plasma expands upwards and fills the newly formed coronal loops. Magnetically detached from the coronal current sheet, the flare loop plasma then cools and as a consequence is gradually observed at soft X-ray (SXR), extreme ultra-violet (EUV), UV and H$\alpha$ wavelengths (\eg, \cite{2006SoPh..234..273V}; for reviews see \cite{lrsp-2008-1} and \cite{2011SSRv..159...19F}).}

{The reconfigured active-region field which maps to the flare ribbons observed in the lower atmosphere is left behind in the form of a shorter (more compact) configuration when compared to that of the pre-flare (pre-CME) state, as implied by the so-called ``coronal implosion'' scenario, originally proposed by \cite{2000ApJ...531L..75H}. Since the frozen-in coronal plasma allows us to trace the structure of the coronal magnetic field indirectly in the form of coronal loops \citep[for a review see][]{2014LRSP...11....4R}, the observable manifestations of such an implosion are more and more frequently reported \citep[see, \eg,][and references therein]{2015A&A...581A...8R}. Most prominent, they are in the form of coronal loops that apparently collapse (shrink) to lower heights. Besides, the (partial) deflation of the coronal magnetic field in the course of eruptive flares is thought to have a detectable impact on the low-atmosphere magnetic field below, \eg, in the form of an increase of the horizontal photospheric field (see \cite{2008ASPC..383..221H} and also, \eg, \cite{2012ApJ...748...77S}). 

Direct measurements of the coronal magnetic field, at a temporal cadence sufficiently high and spatial resolution sufficiently fine to allow for the investigation of eruptive solar phenomena, are not available to date \citep[\eg, review by][]{2009SSRv..144..413C}. Thus, indirect ways to explore their magnetic nature are used, most widely static representations of the 3D coronal magnetic field.} Among the currently popular methods are nonlinear force-free (NLFF) models {\citep[for a review see][]{2014A&ARv..22...78W}}. These methods are based on the vector magnetic field information measured at photospheric levels. They solve a set of simplified ideal MHD equations, in the limit of negligible electric field and vanishing electron diffusivity. Importantly, such methods lack a force exerted by the magnetic field that acts on the charges that produced the magnetic field in first place (\ie, $\jvec\times\Bvec=0$, where $\jvec$ denotes the electric current density and $\Bvec$ is the magnetic field). Such models are valid in a low-$\beta$ environment ($\beta<<1$), where $\beta$ denotes the ratio of gas and magnetic pressure. Assuming the active-region corona to represent such a force-free equilibrium environment, this approach is a valid representation for the quasi-static evolution of active-region coronal magnetic field {\citep[for reviews see][]{2012LRSP....9....5W,2015SSRv..tmp...75W}}. {Note that during flares, plasma of high density and high temperature is created and the correspondent coronal volume represents a high-$\beta$ environment, for which a {\it non}-force-free approximation is desirable. Due to the lack of near real-time implementations of such approaches, however, the alternative is to use a NLFF approximation in order to model the quasi-static evolution of the coronal volume above flaring ARs.}

Given boundary conditions that fulfill the requirements of a force-free equilibrium, NLFF models were shown to perform well \citep{2006SoPh..235..161S,2008SoPh..247..269M}. They, however, encounter difficulties when using an actually measured (non force-free) magnetic field as an input \citep[\eg][]{2008ApJ...675.1637S,2009ApJ...696.1780D}. In particular, different implementations for solving the force-free boundary value problem give different answers on the associated coronal magnetic field structure (and related physical parameters). This differences might be more or less pronounced, depending on the spatial resolution of the input data \citep{2015ApJ...811..107D}. The same method may even result in a different model magnetic field when using input data covering the same area on the solar surface but measured by two different instruments \citep{2012AJ....144...33T,2013ApJ...769...59T}. A careful interpretation of the resulting NLFF model and a critical testing of its quality is therefore essential.

Despite the discussed challenges and the relative simplicity of NLFF models, they are still capable of describing the magnetic nature of the active-region corona, validated by comparison to simultaneously observed coronal emission. Just to name a few, this includes the rearrangement of emerging small-scale twisted magnetic fields in order to establish the (potentially flaring) twisted active-region structure \citep[][]{2012SoPh..278...73V} and the evolution of already emerged twisted active-region fields in the course of strong \citep[\eg,][]{2012ApJ...748...77S} and weak \citep[\eg][]{2014SoPh..289.1153G,2014ApJ...780..102T} flaring activity.

In the present study, we investigate the flaring activity and associated coronal magnetic field evolution of AR NOAA~11261, around a long-duration, eruptive M1.4~flare on 2011 August~2 {\citep[SOL2011-08-02T06:19, following the convention suggested in][]{2010SoPh..263....1L}}. The flare-related phenomena include well-defined flare ribbons, a coronal wave and an Earth-directed (halo) CME. A prominent filament was associated to the northernmost negative sunspot of the AR and the neighboring positive-polarity plage. It was largely unaffected by the flaring activity during August~1--3 (six M-class flares and numerous C-flares) but erupted on August~4, in association with a \goes\ X-class flare. \cite{2014ApJ...785...88Z,2014ApJ...795..175Z} studied in detail the filament eruption on August~4 and suggested that it was caused by a torus instability.

The M1.4~flaring activity on August~2 was not induced from or had an effect on the coronal structures of the ARs neighboring NOAA~11261. The relative isolation from the neighboring ARs, as well as its location relatively close to disk center (situated at heliographic coordinates N17W12 at the peak time of the flare), quantifies NOAA~11261 as a suitable candidate for the purpose of this study: to investigate the temporal and spatial association between the flare-related signatures observed at (extreme) UV and X-ray wavelengths, and the underlying coronal magnetic field (the latter resulting from NLFF magnetic field modeling). We attempt to reconstruct the coronal magnetic field rooted in flaring regions in order to study the magnetic field configuration previously established by magnetic reconnection, ultimately allowing us to draw conclusions on the upward motion of the current sheet's lower tip in the corona.

\section{Data and Methods}

\subsection{Imaging data and flare pixel detection}

Within this work, we present the analysis of co-temporal (extreme) UV emission, observed with the {\it Solar Dynamics Observatory} \citep[\sdo;][]{2012SoPh..275....3P} Atmospheric Imaging Assembly \citep[AIA;][]{2012SoPh..275...17L}, the Extreme UltraViolet Imager \citep[EUVI;][]{2008SSRv..136...67H} on board the {\it Solar Terrestrial Relations Observatory} \citep[\stereo;][]{2008SSRv..136....5K}, as well as of X-ray emission, measured by the {\it Reuven Ramaty High Energy Solar Spectroscopic Imager} \citep[\rhessi;][]{2002SoPh..210....3L}.

In particular, we use the AIA 1600\,\AA\ (UV) passband which is sensitive to plasma at temperatures around 5000\,K (photosphere) but also transmits emission from two transition-region C~{\sc iv} lines, forming at about 0.1\,MK. We use images at a 1-min cadence and with a spatial resolution of $1\farcs2$ in order to trace the evolution of the flare ribbons. We apply a 3-min running-median filter in order to eliminate transient features which are not related to the flaring activity. The flare ribbons, visible in the form of ridges of enhanced emission in the UV images, are then traced by an automated method. This method marks pixel locations as flare-related whenever its intensity value exceeds the {98 {\it percentile}} of the entire data series. Thus, only the brightest pixels, based on the relative occurrence of intensity values, are marked as flare-related. Additionally, we use AIA 131\,\AA\ (EUV) images sensing emission from Fe~{\sc viii} and Fe~{\sc xxi}, with peak formation temperatures at $\log T=5.6$ (transition region) and $\log T=7.0$ (flaring plasma), in order to trace the hot flare plasma.

In order to locate flare-associated X-ray sources, we use \rhessi\ images with a spatial resolution of $2\farcs3$, reconstructed using the Clean algorithm \citep{2002SoPh..210...61H} for different energy bands, spanning the range 3--50\,keV. \rhessi\ coverage was not 100\% during the entire flare investigated here. It covered part of the early (rising) phase of the flare (from 04:28~UT to 05:28~UT) and the peak and early declining phase (from 06:04~ UT to 06:58~UT).

The advantageous position of the \stereo-A spacecraft during the event under study, in almost perfect quadrature view (separation angle with Sun--Earth line $\approx$\,100$^\circ$), allowed us to use EUVI 195\,\AA\ images for the purpose of observing the post-flare loop arcade above the \stereo-A limb. The light at 195\,\AA\ is primarily emitted by Fe~{\sc xii} and corresponds to plasma temperatures of 1.4~MK, \ie, to the signatures of the cooled flare loops' plasma. We use images at a 5-min cadence with a spatial resolution of $\approx$\,$3\farcs2$.

All images are prepared and co-registered using standard IDL mapping software, where we de-rotated all data to the peak time of the flare (06:19~UT) in order to account for the effects of differential rotation.

\begin{figure}
	\centering
	\includegraphics[width=\columnwidth]{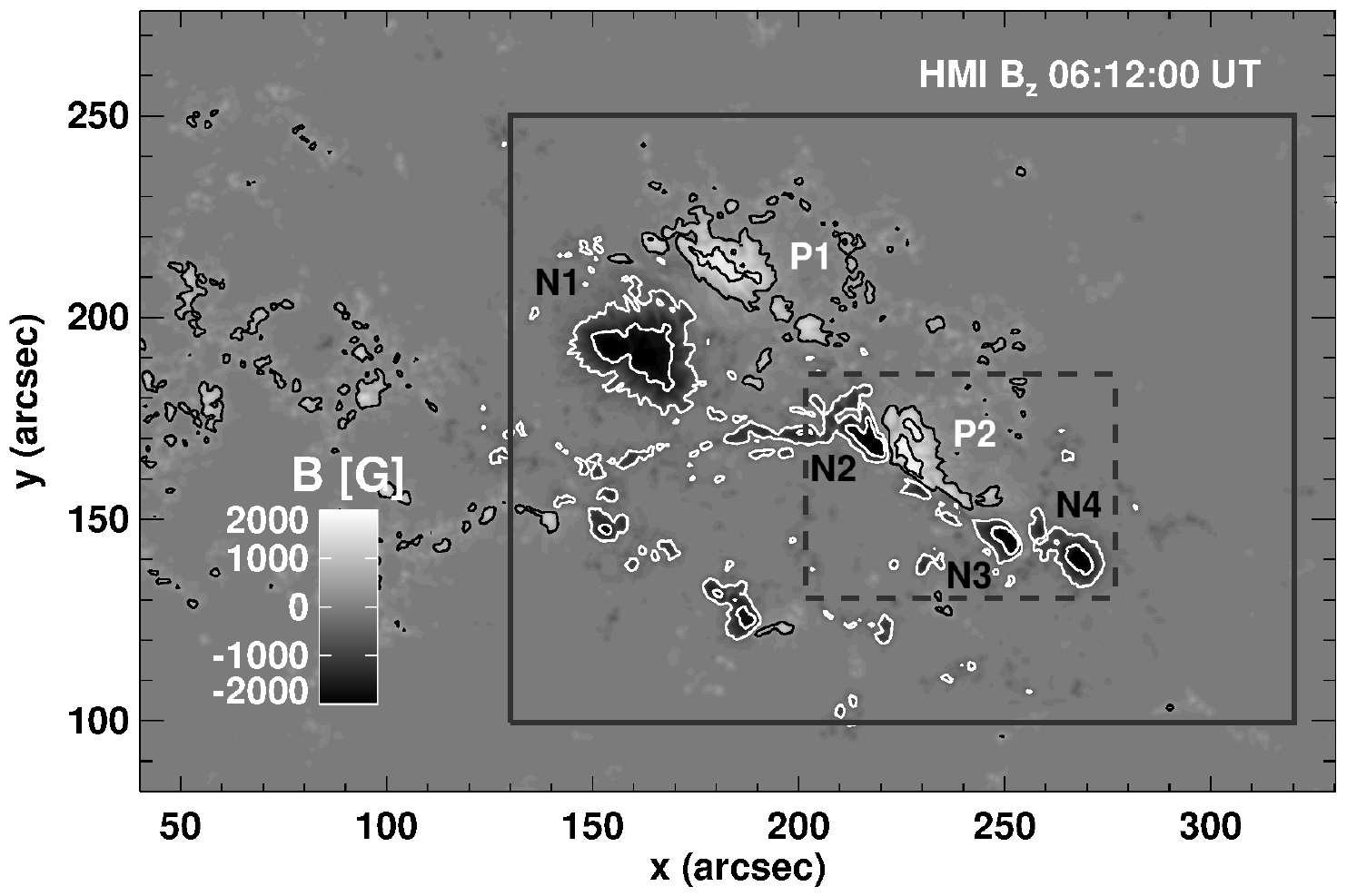}
	\caption{Vertical magnetic field on August~2 at 06:12~UT (gray-scale background scaled to $\pm$2\,kG, with black/white colors representing negative/positive polarity). Black (white) contours outline positive (negative) vertical magnetic field of $\pm$[0.5,1.5]\,kG. Labels N1--N4 mark the negative polarity patches associated to four observed sunspots, while P1 and P2 mark the associated positive polarity patches. The solid (dashed) rectangle marks the sub-field used for description of the flare-associated emission (pre-flare magnetic field evolution). Units are arc-seconds from Sun center. (A color version of this figure is available in the online journal.)\\}
	\label{fig:fig1}
\end{figure}

\subsection{Magnetic field modeling}

In order to interpret the flare-related signatures seen in coronal (extreme) UV images and to establish a link to the observed X-ray flare emission, we employ NLFF coronal magnetic field models using \sdo\ Helioseismic and Magnetic Imager \citep[HMI;][]{2012SoPh..275..229S,2014SoPh..289.3483H} vector magnetic field data. We use an optimization algorithm that seeks the force-free and solenoidal magnetic field configuration in a 3D cubic domain, given specified boundary conditions. The algorithm is halted once an approximately steady state of minimal Lorentz force and solenoidality is found. 

We use nearly flux-balanced (flux-imbalance $\lesssim$\,3\%) HMI vector magnetic field maps with a 12-min cadence {and a spatial resolution of $\approx$\,$1\farcs0$}. We de-rotate all magnetic field maps to the peak time of the flare, using standard IDL mapping software and project the measured (image-plane) magnetic field vector to a local (heliographic) coordinate system, in order to minimize projection effects \citep[following][]{1990SoPh..126...21G}. In a second step, the de-projected non force-free photospheric magnetic field vector data is driven to a more force-free consistent configuration \citep[simultaneously preserving agreement with the measured data;][]{2006SoPh..233..215W}. This improved boundary data is then supplied to the optimization scheme as a lower boundary condition. Using a grid refinement scheme and simultaneously accounting for measurement uncertainties, the optimization method delivers a near force-free and near solenoidal magnetic field solution \citep[for details of the method see][and Sect.\ 2.2.1 of \cite{2015ApJ...811..107D}]{2010A&A...516A.107W,2012SoPh..281...37W}. 

The extension of the computational domain is $290\farcs4\times193\farcs7\times129\farcs0$ ($\approx$\,$213.9\times142.7\times95.0$\,Mm), centered around ($185\farcs2$,$179\farcs2$) from Sun center. The photospheric area covered in the photospheric HMI vector maps that are used as input to the NLFF modeling is shown in Figure~\ref{fig:fig1}. The solid and dashed rectangles outline the sub-field used for analysis of the flare-associated emission and pre- (early) flare development, respectively.

In order to quantify the goodness of the obtained NLFF model solutions, we employ controlling metrics that have been put forward for measuring the relative success of NLFF modeling: the current-weighted average of the sine of the angle between the model magnetic field and the electric current density, as well as the volume-averaged fractional flux \citep[$\sigma_\jvec$ and $\langle|f_i|\rangle$, respectively; see][]{2000ApJ...540.1150W}. While the former tests the success of recovering a force-free solution, the latter quantifies how close the final state is to solenoidality. For a perfectly force-free and solenoidal solution, $\sigma_\jvec=0$ (\ie, $\jvec\times\Bvec=\mathbf 0$ since $\Bvec\parallel\jvec$) and $\langle|f_i|\rangle=0$ (\ie, $\nabla\cdot\Bvec=0$), respectively. In general, NLFF solutions based on real data deviate from a perfectly force-free and solenoidal state, which results in values $0<[\sigma_\jvec,\langle|f_i|\rangle]\leq1.0$. The metrics of the entire series of NLFF models discussed in the present study (between 05:00~UT and 07:00~UT with a 12-min time cadence) are $\sigma_\jvec=\mathcal{O}(10^{-1})$ and $\langle|f_i|\rangle=\mathcal{O}(10^{-4})$, showing that our NLFF models may be considered as to be qualified for the use within the presented study.

\section{Results}

\subsection{Pre-flare configuration and development}\label{sss:preflare_evol}

\begin{figure}
	\centering
	\includegraphics[width=0.75\columnwidth]{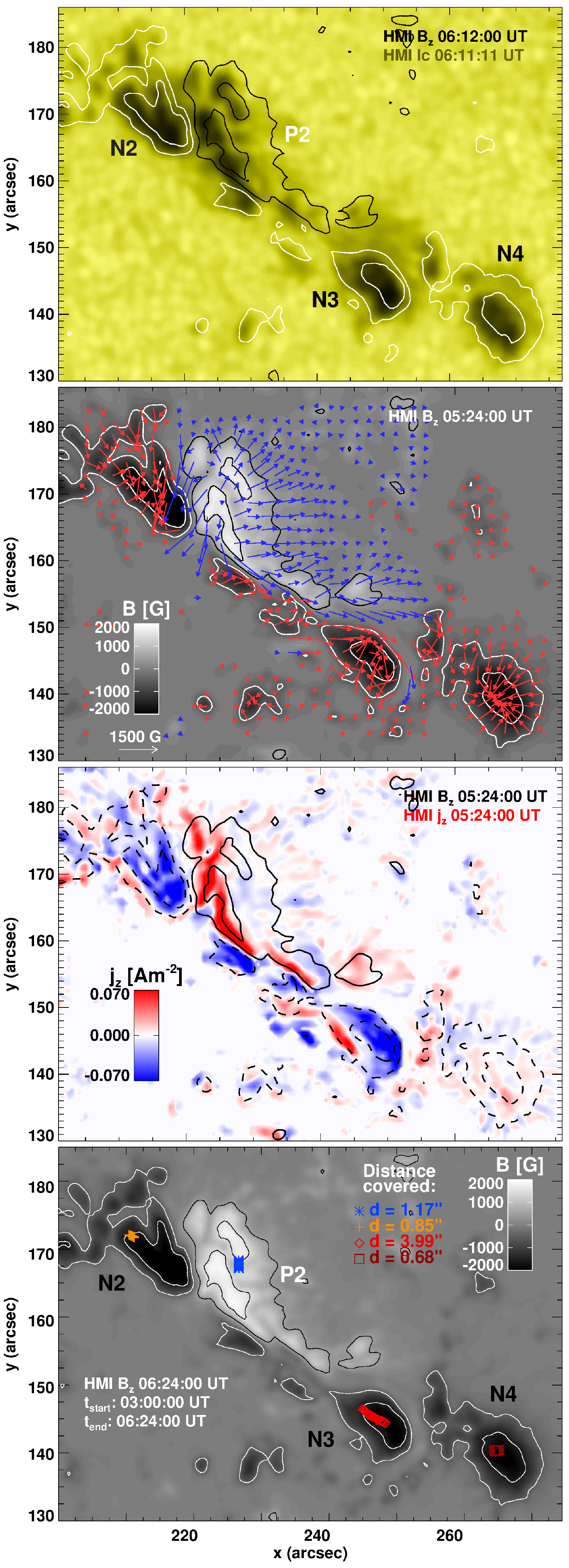}
	\put(-194,493){\Large\sf\black (a)}
	\put(-194,370){\Large\sf\black (b)}
	\put(-194,248){\Large\sf\black (c)}
	\put(-194,125){\Large\sf\black (d)}
	\caption{(a) HMI continuum image showing the $\delta$-spot configuration in the center of NOAA~11261. Black/white contours outline the $\pm$[0.5,1.5]\,kG levels of positive/negative vertical magnetic field $\Bz$. Labels N2 and P2 mark the mixed polarity associated to the $\delta$-spot. N3 and N4 mark the negative polarity associated to two further sunspots. (b) Orientation and strength (proportional to the length of the arrows shown) of the horizontal magnetic field $\Bh$ (blue/red color correspond to $\Bh$ being associated to positive/negative $\Bz$). $\Bz$ at 06:12~UT is shown as gray-scale background (scaled to $\pm$2\,kG), where white and black color represents positive and negative polarity, respectively (contours as in (a)). (c) Co-temporal positive (red) and negative (blue) vertical electric current density, $\jz$, calculated from $\Bh$ and scaled to $\pm0.07$\,A\,m$^{-2}$ (calculated only where $|\Bz|$\,$>$\,1\,G and $\Bh$\,$>$\,10\,G). Solid and dashed lines correspond to black and white contours in the other panels, respectively. (d) Motion of flux-weighted polarity centers N2 (yellow crosses), P2 (blue stars), N3 (light red triangles), and N4 (dark red squares). $\Bz$ at 04:48~UT is shown as gray-scale background. The start and end time of tracking the polarity-center motion is indicated at the bottom left corner. Black/white contours outline the $\pm$[0.5,1.5]\,kG levels of positive/negative $\Bz$. The FOV corresponds to the dashed outline in Figure~\ref{fig:fig1}. Units are arcseconds from Sun center. (A color version of this figure is available in the online journal.)}
	\label{fig:fig2}
\end{figure}
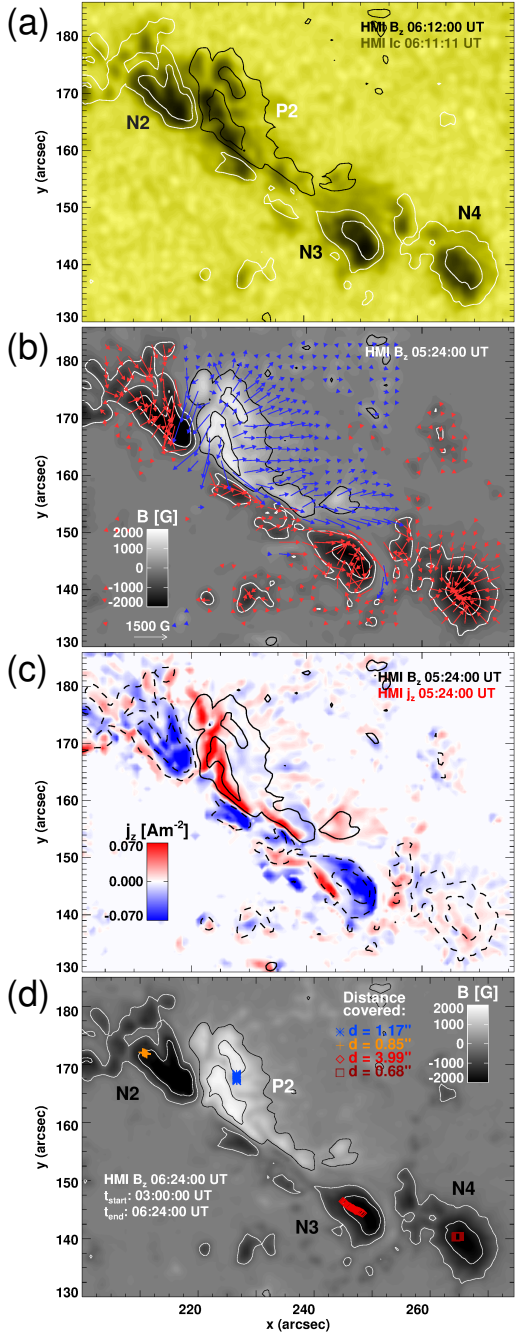

The reason for the ongoing moderate to strong flaring activity during 2011 August 1--4 was rooted in the complex magnetic structure of the NOAA~11261, a $\beta\gamma\delta$-configuration. The AR was composed of four sunspots on August~2, three of which were dominated by a negative magnetic polarity (marked as N1/P1, N3 and N4 in Figure~\ref{fig:fig1}). A less compact system of mixed negative and positive polarities (N2/P2), formed a fourth sunspot, a $\delta$-spot (spatially separated umbrae of opposite polarity located within a single penumbra; see also Figure~\ref{fig:fig2}a). Note that the AR evolved into a three-sunspot system within the next $\approx$\,24~hours. While the sunspots associated to N1/P1 and N2/P2 kept their identities, N3 and N4 merged {to form a larger sunspot}.

In Figure~\ref{fig:fig2}b--d, we {take a closer look at} the photospheric magnetic field configuration in the AR center prior to the M1.4~flare. Highly sheared fields (indicated by arrows in Figure~\ref{fig:fig2}b) are found in two places: to the {south-west} of the positive polarity P2 (which, together with N2 forms the $\delta$-spot) and to the west of the negative polarity N3 (towards the negative polarity patch N4). Note that highly sheared magnetic fields (not necessarily together with a $\delta$-spot) have often been identified as a fruitful environment for flaring activity \citep[\eg][and references therein]{2009SSRv..144..351V}. Not surprisingly, the strongest vertical electric current densities are found in those places where the magnetic shear is high (compare Figure~\ref{fig:fig2}c). 

In order to investigate the reasons for the strong electric current concentrations prior to the M1.4~flare, we track the motion of the individual polarity patches' flux-weighted center with time. We find that N3 (marked as red triangles in Figure~\ref{fig:fig2}d) travels a significant distance in south-western direction until shortly the flare peak ($d\approx4\farcs0$ in the period 03:00--06:24~UT, \ie~with a mean velocity of $\approx$\,0.24\,km\,s$^{-1}$). This naturally resulted in an enhanced shear of the field connecting N3 and its surrounding (represented by arrows in Figure~\ref{fig:fig2}b). In contrast, none of the other polarity centers showed such a clear motion (stars, crosses and squares in Figure~\ref{fig:fig2}d refer to the flux-weighted polarity centers of P2, N2, and N4, respectively). For completeness, we note that N3 continued its journey towards the south-west of the AR and merged with N4 on August~3, when positive fields in between were canceled \citep[see][]{2014ApJ...785...88Z}.

\subsection{Flare-associated emission}

\subsubsection{Temporal evolution of (E)UV and X-ray emission}\label{sss:goes}

\begin{figure}
	\centering
	\includegraphics[width=\columnwidth]{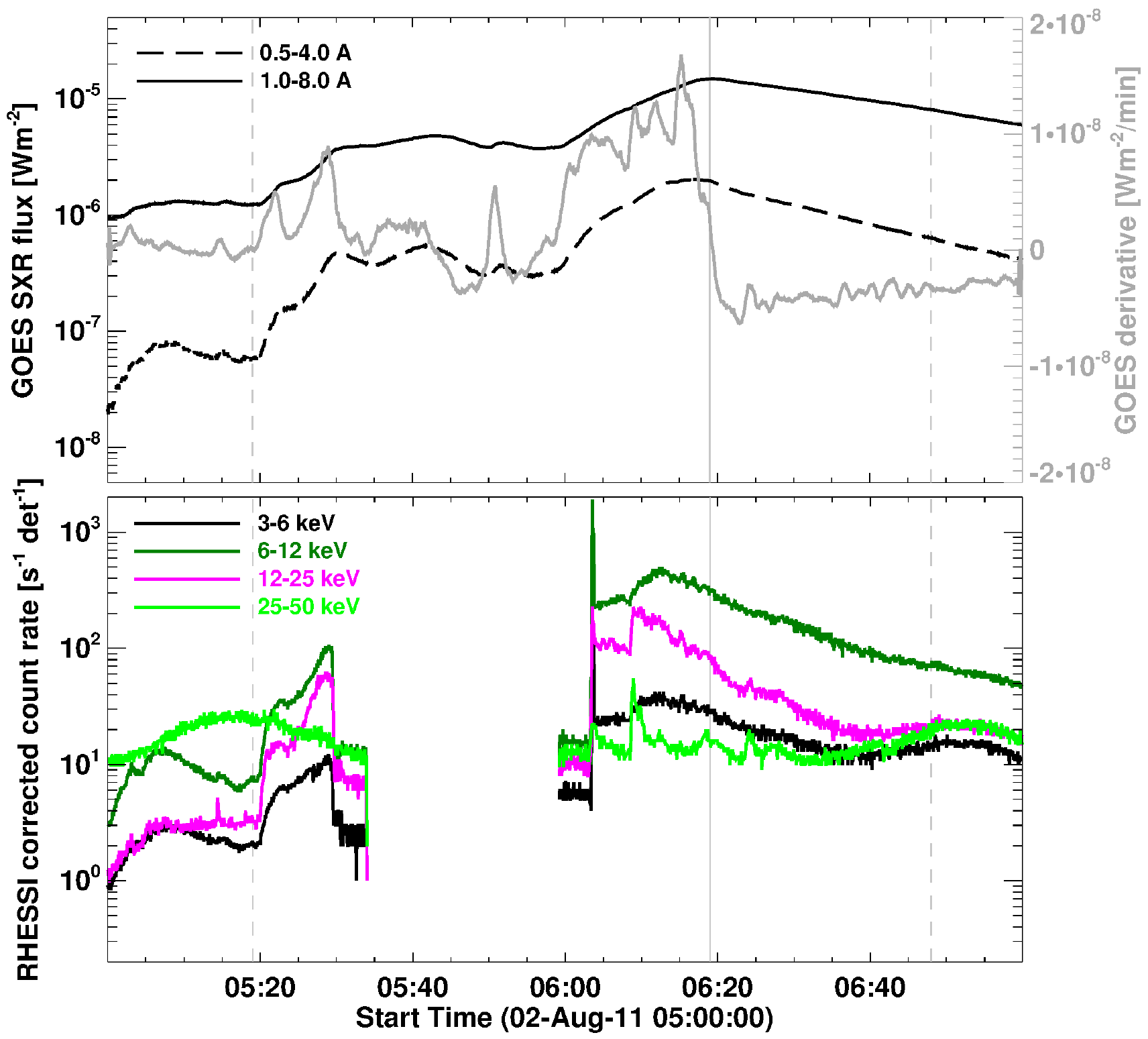}
	\put(-50,205){\Large\sf\black (a)}
	\put(-50,102){\Large\sf\black (b)}
	\caption{(a) \goes\ SXR flux in the 1.0--8.0\,\AA\ (solid black curve) and 0.5--4.0\,\AA\ (dashed black curve) wavelength band from 05:00 to 07:00~UT. The gray solid curve shows the derivative of the \goes\ SXR flux. (b) \rhessi\ X-ray counts (corrected) in the 6--12\,keV (black), 12--25\,keV (dark green), 25--50\,keV (magenta), and 25--50\,keV (light green) energy band. The vertical dashed lines mark the nominal start (05:19~UT) and end time (06:48~UT) of the M1.4~flare, based on the \goes\ SXR flux. The solid vertical line marks the flare peak time (06:12~UT). Horizontal arrows indicate two distinct phases during the impulsive phase. (A color version of this figure is available in the online journal.)\\}
	\label{fig:fig3}
\end{figure}

The {SXR emission associated to SOL2011-08-02T06:19} starts at 05:19~UT and peaks at 06:19~UT (Figure~\ref{fig:fig3}a). The SXR emission shows a complex rise-and-fall pattern towards the peak emission. Starting from around 05:19~UT, the SXR flux rises quickly and reaches C3-level ($\approx$\,$3\times10^{-5}$\,W\,m$^{-2}$) at $\approx$\,05:30~UT (phase {\sc I}). It fluctuates around that level until $\approx$\,06:00~UT. This is followed by a further rise until 06:19~UT (phase {\sc II}) when the peak flux of $1.4\times10^{-6}$\,W\,m$^{-2}$ is reached. Afterwards, the SXR activity level does not drop to background B-level before $\sim$11:00~UT, classifying the event as a long-duration event.

Since \rhessi\ missed part of the impulsive phase, we also plot in Figure~\ref{fig:fig3}a the derivative of the \goes\ 1.0--8.0\,\AA\ flux, which serves as a proxy for the evolution of the energy in flare-accelerated electrons \citep[Neupert effect; see, \eg,][]{1993SoPh..146..177D,2002A&A...392..699V}. The thermal 6--12\,keV and 12--25\,keV count rate shows distinct peaks at the beginning of phase {\sc I} (rise to C3-level) and during large parts of phase {\sc II} (rise to M1-level). The non-thermal (25--50\,keV) emission shows only one distinct peak in phase {\sc II}, around 06:08~UT (light green curve in Figure~\ref{fig:fig3}b).

\begin{figure*}
	\epsscale{1.0}
	\centering
	\includegraphics[width=2.0\columnwidth]{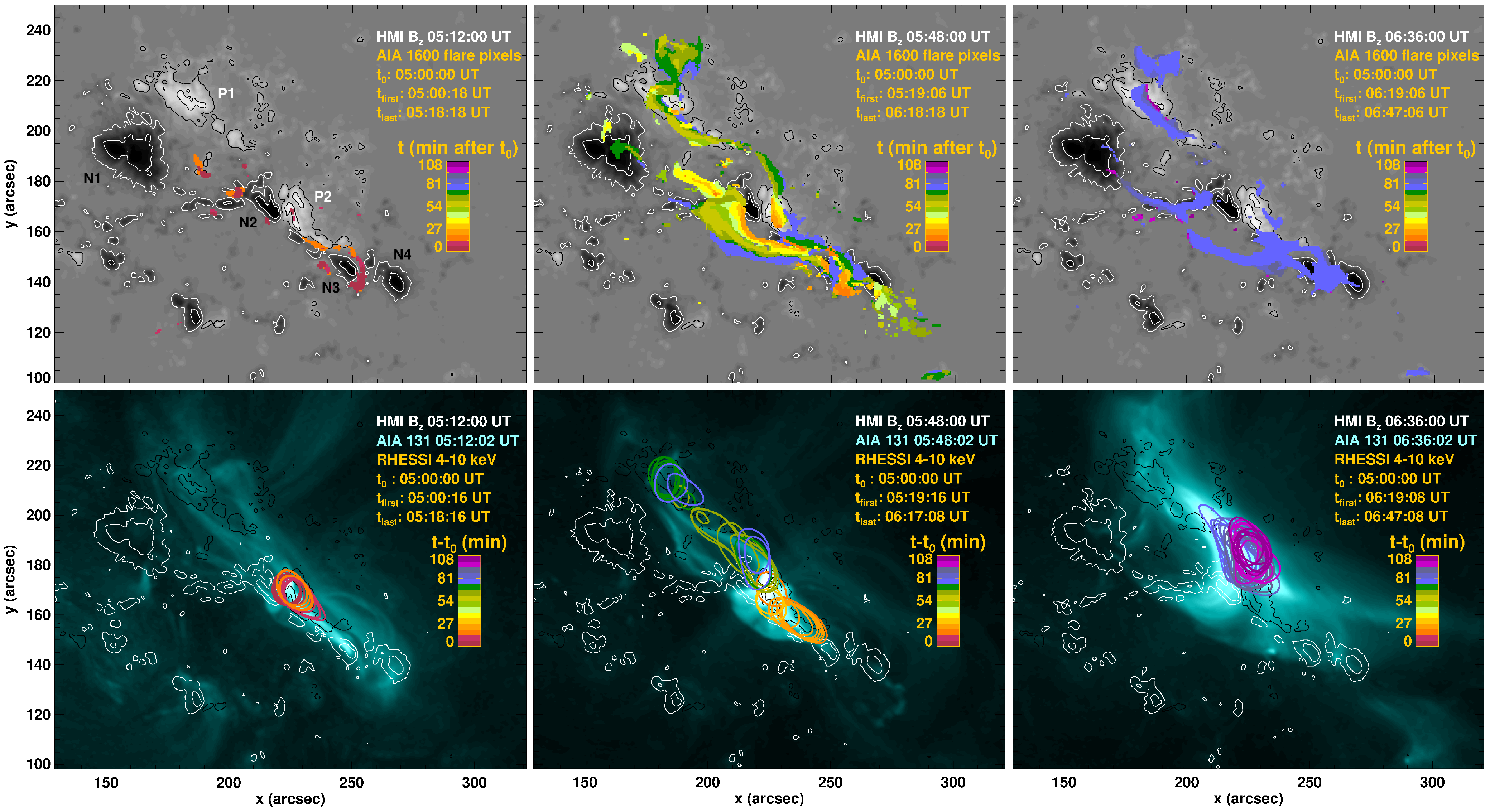}
	\put(-468,252){\Large\sf\white (a)}
	\put(-308,252){\Large\sf\white (b)}
	\put(-150,252){\Large\sf\white (c)}
	\put(-468,124){\Large\sf\white (d)}
	\put(-308,124){\Large\sf\white (e)}
	\put(-150,124){\Large\sf\white (f)}
	\caption{{\it Top panels:} Locations attributed to activity in the (a) pre-flare (05:00--05:19~UT), (b) impulsive (05:19--06:19~UT), and (c) decay (06:19--06:48~UT) phase of the M1.4~flare, based on automated tracking of flare pixels in the AIA 1600\,\AA\ images. The flare pixels are color-coded according to their time of first appearance (minutes after the reference time $t_0$=05:00~UT). The times of the first and last detected flaring pixel within the distinct phases are indicated as $t_{\rm first}$ and $t_{\rm last}$ in the upper right corners, respectively. The gray-scale background resembles $\Bz$ at a time indicated in the top right corner of each panel. Black/white contours outline the $\pm[0.5,1.5]$\,kG levels of $\Bz$. Labels N1--N4 in panel (a) mark the negative polarity patches associated to four observed sunspots, while P1 and P2 mark the associated positive polartiy patches. {\it Bottom panels:} AIA 131\,\AA\ images (co-temporal to the $\Bz$ shown in the top row), together with \rhessi\ 4--10\,keV image contours. Color-coded contours are drawn at 75\% of the maximum intensity at each time instance. The color-coding resembles the progression of time (minutes after the reference time $t_0$). Black/white contours outline $\Bz=\pm[0.5,1.5]$\,kG, respectively. The FOV corresponds to the solid outline in Figure~\ref{fig:fig1}. Units are arc-seconds from Sun center. (A color version of this figure is available in the online journal.)\\}
	\label{fig:fig4}
\end{figure*}

\subsubsection{Spatial distribution of (E)UV and X-ray emission}

We inspect the evolution of the flare ribbons, traced at UV wavelengths (using AIA 1600\,\AA\ images) during different evolutionary steps: the pre- (early) flare phase between 05:00~UT and 05:19~UT (ending with the nominal start time of the flare; see Figure~\ref{fig:fig4}a), the impulsive phase from 05:19~UT to 06:19~UT (ending with the peak time of the flare; see Figure~\ref{fig:fig4}b), and the decay phase starting from the peak time, ending with the nominal end time of the flare (06:48~UT; see Figure~\ref{fig:fig4}c). In order to visualize the progression in time, we color-code the detected flare pixels according to the time when they were associated to flaring emission for the first time.

We notice only localized kernels of enhanced emission during the pre- (early) flare phase especially to the west of N3 (Figure~\ref{fig:fig4}a). Proper ribbons form during the impulsive phase. In this period, the flare ribbons are characterized by a growth in length and, depending on the position along the ribbons, a more or less pronounced increase in separation (Figure~\ref{fig:fig4}b). Note that the southern ribbon intrudes the main negative-polarity sunspot N1. During the declining phase of the flare, the ribbons further increase their separation but do not grow in length anymore (Figure~\ref{fig:fig4}c). 

Inspection of the co-temporal thermal \rhessi\ X-ray sources in the 4--10\,keV energy band reveals a localized narrow source, on top of the already enhanced pre- (early) flare AIA 131\,\AA\ emission (Figure~\ref{fig:fig4}d). It is associated to the mixed polarity system P2/N2, in contrast to the observed AIA 1600\,\AA\ pre- (early) flare emission, which mainly formed co-spatial with the region of strong shear in the south-west of the AR (compare Figure~\ref{fig:fig4}a). 

During the impulsive phase of the flare, the temporal evolution of thermal X-ray emission follows the ridge of bright AIA 131\,\AA\ emission to the north-east, towards the main positive-polarity spot P1 (Figure~\ref{fig:fig4}e). It follows the trace of the northern AIA 1600\,\AA\ flare ribbon (compare Figure~\ref{fig:fig4}b). Note that during the early impulsive phase (05:19--05:30~UT) the thermal X-ray sources are found only around N2/P2 and in the south-west of it (orange and yellow contours in Figure~\ref{fig:fig4}e). Only during the late impulsive phase, they progress towards the north-east (towards P1; green and blue contours in Figure~\ref{fig:fig4}e). Note that this is in temporal agreement with the two phases identified during the impulsive phase from the \goes\ SXR light curve (see Sect.\ \ref{sss:goes}).

The decay phase is characterized by a localized source on top of the post-flare loops seen in AIA 131\,\AA (Figure~\ref{fig:fig4}f) and bridge the locations associated to AIA 1600\,\AA\ flare ribbon emission in the atmosphere below (compare Figure~\ref{fig:fig4}c). It apparently shrinks in size and outlines the apexes of the observed post-flare loop system.

\begin{figure*}
	\epsscale{1.0}
	\centering
	\includegraphics[width=2.0\columnwidth]{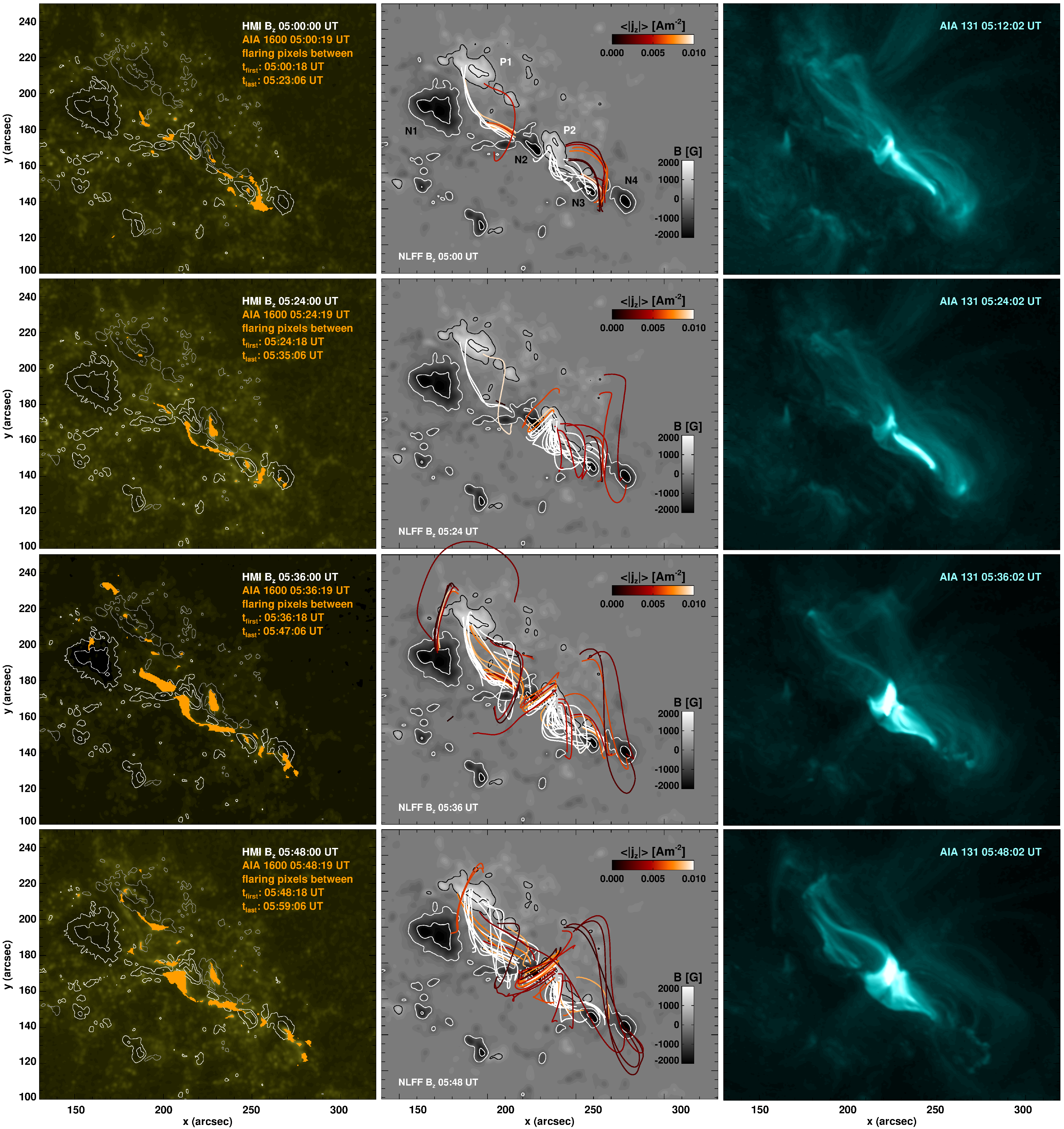}
	\put(-468,505){\Large\sf\white (a)}
	\put(-308,505){\Large\sf\white (b)}
	\put(-150,505){\Large\sf\white (c)}
	\put(-468,378){\Large\sf\white (d)}
	\put(-308,378){\Large\sf\white (e)}
	\put(-150,378){\Large\sf\white (f)}
	\put(-468,251){\Large\sf\white (g)}
	\put(-308,251){\Large\sf\white (h)}
	\put(-150,251){\Large\sf\white (i)}
	\put(-468,124){\Large\sf\white (j)}
	\put(-308,124){\Large\sf\white (k)}
	\put(-150,124){\Large\sf\white (m)}
	\caption{{Spatio-temporal evolution of the flare kernels (ribbons) and magnetic fields during the pre-flare and early impulsive phase.} {\it Left panels:} Accumulated flare pixel positions, recorded within time ranges indicated at the top right of each panel (orange filled contours) on top of the AIA 1600\,\AA\ images at the beginning of the time interval. Black and white contours outline the co-temporal $\pm$[0.5,1.5]\,kG of positive and negative $\Bz$, respectively. {\it Middle panels:} The gray-scale background resembles $\Bz$ on the NLFF lower boundary, co-temporal to the AIA 1600\,\AA\ images and scaled to $\pm$2\,kG. NLFF field lines that originate from the accumulated flare pixels are shown on top. The coloring of the field lines is given by the mean absolute vertical current density, $\langle|\jz|\rangle$, at both field line's footpoints and their eight nearest neighbors. We show only field lines that originate from locations where {$\Bz$\,$>$\,50\,G} and which close within the AR center. {\it Right panels:} AIA 131\,\AA\ images, showing hot flare plasma, co-temporal to the AIA 1600\,\AA\ images. The FOV corresponds to the solid outline in Figure~\ref{fig:fig1}. Units are arc-seconds from Sun center. (A color version of this figure is available in the online journal.)\\}
	\label{fig:fig5}
\end{figure*}

\begin{figure*}
	\epsscale{1.0}
	\centering
	\includegraphics[width=2.0\columnwidth]{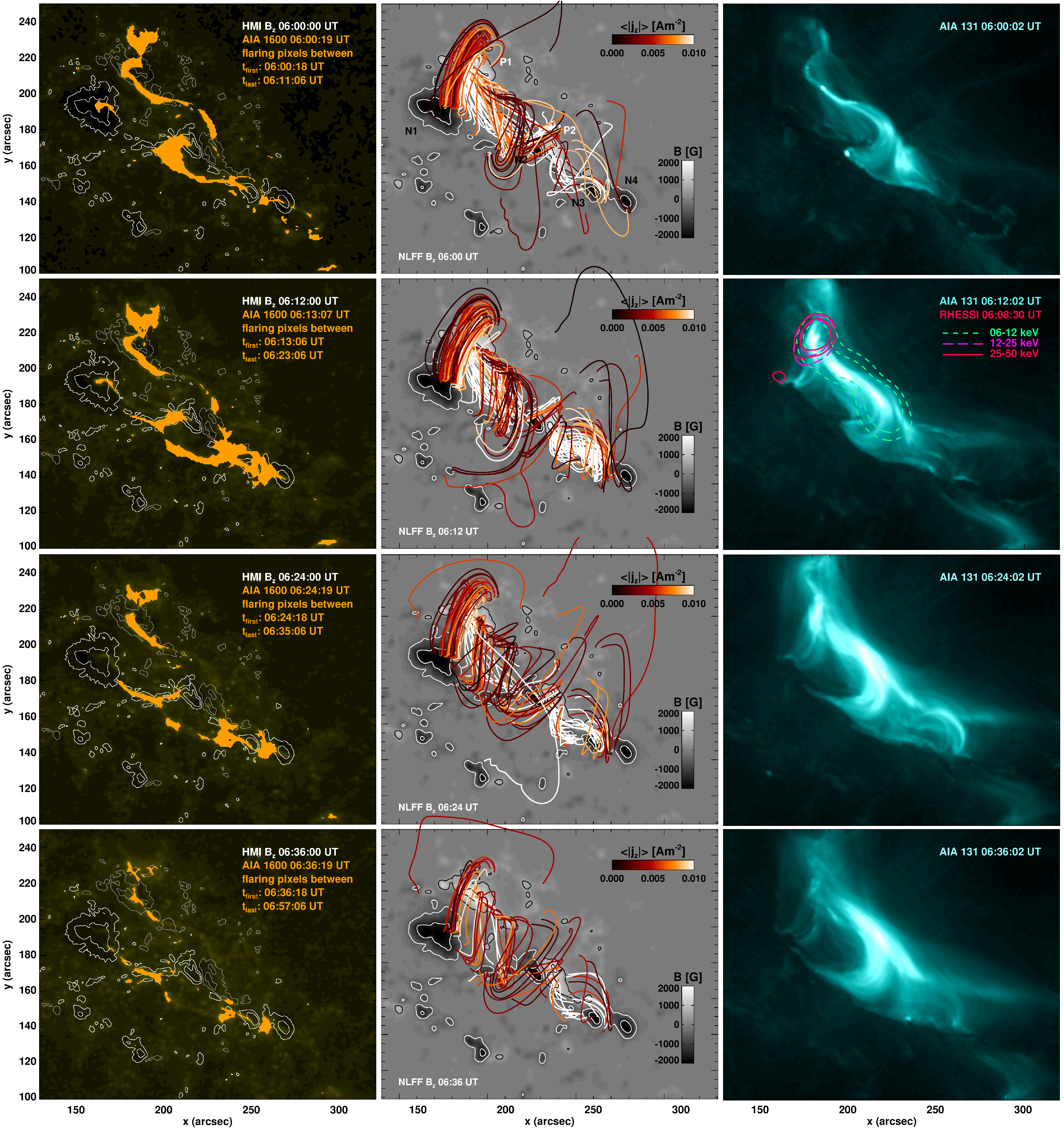}
	\put(-468,505){\Large\sf\white (a)}
	\put(-308,505){\Large\sf\white (b)}
	\put(-150,505){\Large\sf\white (c)}
	\put(-468,378){\Large\sf\white (d)}
	\put(-308,378){\Large\sf\white (e)}
	\put(-150,378){\Large\sf\white (f)}
	\put(-468,251){\Large\sf\white (g)}
	\put(-308,251){\Large\sf\white (h)}
	\put(-150,251){\Large\sf\white (i)}
	\put(-468,124){\Large\sf\white (j)}
	\put(-308,124){\Large\sf\white (k)}
	\put(-150,124){\Large\sf\white (m)}
	\caption{{Same as in Fig.~\ref{fig:fig5} but for the late impulsive and decay phase.} (A color version of this figure is available in the online journal.)\\}
	\label{fig:fig6}
\end{figure*}

\subsection{Flare-related coronal magnetic field}

\subsubsection{Spatio-temporal evolution}\label{sss:mf_config}

After identifying the locations associated to flaring emission in the low atmosphere, we are able to inspect the associated coronal (post-reconnection) magnetic field. To this aim, we consider the accumulated flare pixel positions detected during specific time intervals. We use 24-min intervals for the pre- (early) flare (Figure~\ref{fig:fig5}a) and late decay (Figure~\ref{fig:fig6}j) phase. During the rest of the flare, we use 12-min intervals, which is dictated by the time cadence of our NLFF models. The according start and end times are indicated at the top right corner of the left panels in Figures~\ref{fig:fig5} and \ref{fig:fig6}.

We use the flare pixels as start locations to calculate field lines from the NLFF model at the beginning of each considered time interval. We do so because we assume that any observed chromospheric and coronal (flare) emission is due to the reconfiguration of the magnetic field prior to the observation of the (flare) emission itself. {Importantly, the field lines deduced in that way cannot be regarded as a visualization of the instantaneous reconfiguration due to magnetic reconnection but rather an overall impression of the field that has been involved during the considered time interval.} A subset of these calculated model field lines is shown in the middle panels of Figures~\ref{fig:fig5} and \ref{fig:fig6}. {Field lines are traced only from locations with a vertical magnetic field magnitude $>$\,50\,G and} are color-coded according to the mean vertical electric current density, $\langle|\jz|\rangle$, at both field line footpoints and their eight nearest neighbors. Note that we show a subset of field lines only in order to present the underlying coronal magnetic field configuration in a clear manner. The findings discussed in the following also hold for the full set of model field lines emanating from flaring pixels.

On overall, we note a striking similarity of the observed structures in the AIA 131\,\AA\ images and the NLFF model field lines calculated from the flaring pixels, especially from the pre- (early) flare phase until the late impulsive phase (compare Figure~\ref{fig:fig5}c,e,h,k to Figure~\ref{fig:fig5}d,f,i,m, respectively). In particular, field lines originating from places of strong $\langle|\jz|\rangle$ (white colored lines) are often found co-spatial with bright AIA 131\,\AA\ emission, suggesting enhanced radiation due to the dissipation of strong electric currents. A system of low-lying, highly sheared magnetic field lines connects P2 and N3 in the pre- (early) flare phase (Figure~\ref{fig:fig5}b). These field lines are co-spatial to bright connections between P2 and N3 seen in AIA 131\,\AA\ images (Figure~\ref{fig:fig5}c). 

{The temporal evolution of the post-reconnection model magnetic field outlines the basic mechanism of eruptive flares. Magnetic reconnection occurs first in the highly sheared and/or twisted core region (close to the main PIL; Figure~\ref{fig:fig5}) and gradually involves the over-arching (more potential) field during later stages of the flare (Figure~\ref{fig:fig6}). For completeness, we note that this conclusion does not depend on using the NLFF models at the beginning of the considered time intervals for field line calculation. One would get the same general picture when performing the field line calculation from, \eg, a single pre-flare NLFF model.}

During the early impulsive phase (05:19--05:30~UT, phase {\sc I} in Figure~\ref{fig:fig3}a), the reconfiguration of the field connecting P2 and its negative-polarity neighborhood dominates the emission picture (Figure~\ref{fig:fig5}c, \ref{fig:fig5}f, and \ref{fig:fig5}i). Only during the late impulsive phase (05:30--06:19~UT, phase {\sc II} in Figure~\ref{fig:fig3}a), also more and more connections to the north-east of the AR (to the positive polarity P1) show clearly enhanced emission (Figure~\ref{fig:fig5}m, Figure~\ref{fig:fig6}c,f), as do the corresponding connectivities in the NLFF models (compare Figure~\ref{fig:fig5}k, Figure~\ref{fig:fig6}b,e, respectively). The single non-thermal X-ray emission peak, discussed in Sect.~\ref{sss:goes}, apparently originated from the flaring loops connecting the two main polarities P1 and N1, as two footpoints are clearly discernible from the \rhessi\ 25--50\,keV images (red solid contours in Figure~\ref{fig:fig6}f). The spatial distribution of the thermal emission (6--12\,keV; green dashed contours in Figure~\ref{fig:fig6}f), on the other hand, outlines the bulk of bright flaring loops connecting N1/P1 to N2/P2.

During the decay phase, we find emission from a growing post-flare loop system (Figure~\ref{fig:fig6}i,m), again well resembled by the NLFF model field lines (compare Figures~\ref{fig:fig6}h,k, respectively). The latter appear to connect locations further and further apart from each other, owing to little shear and/or twist, and consequently connecting regions of low $\langle|\jz|\rangle$. The bright emission can therefore be attributed to plasma which got heated during the flaring process an now looses energy due to the cooling of the plasma to lower temperatures.

\begin{figure*}
	\epsscale{1.0}
	\centering
	\includegraphics[width=2.0\columnwidth]{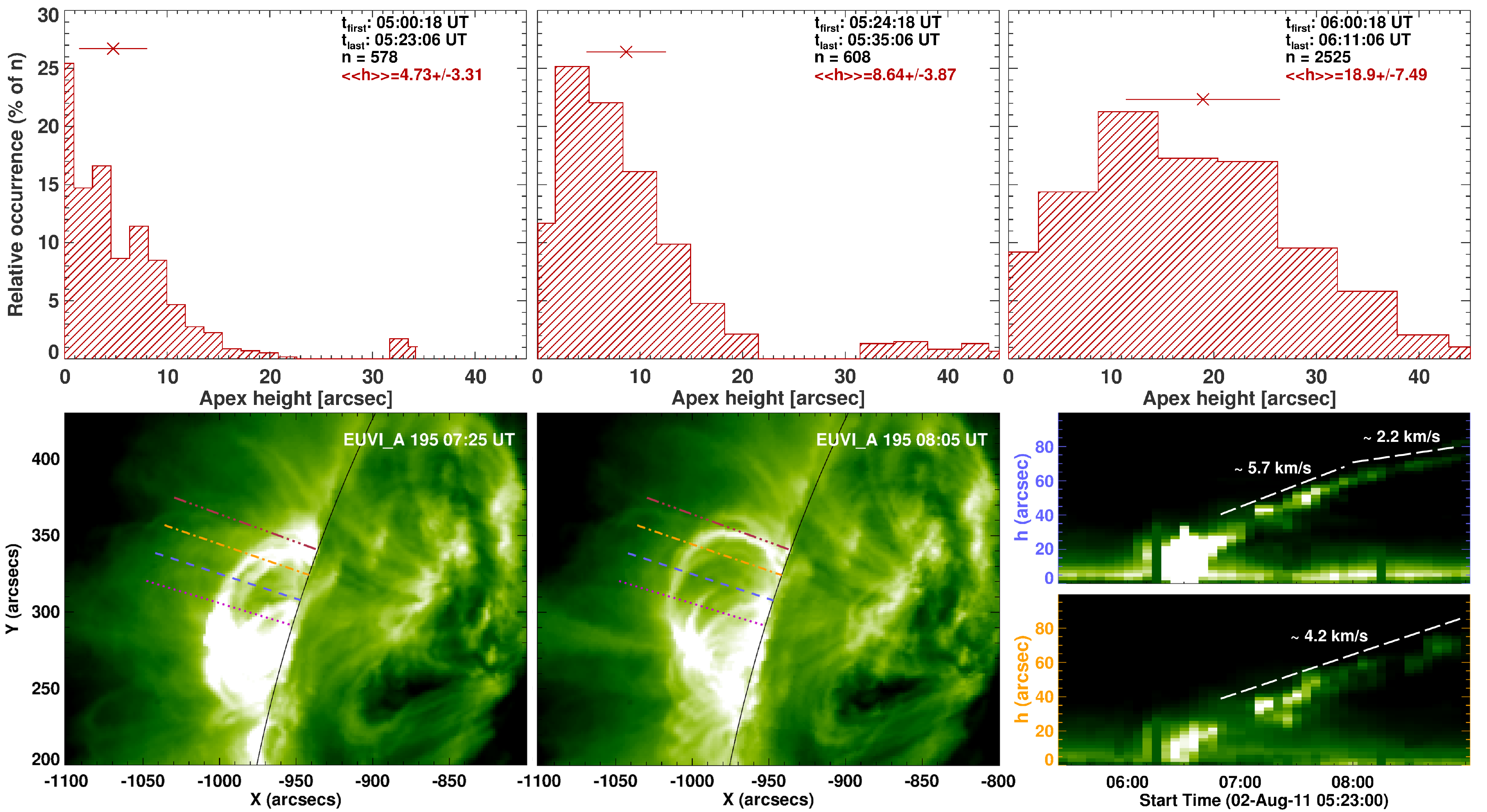}
	\put(-336,252){\Large\sf\black (a)}
	\put(-180,252){\Large\sf\black (b)}
	\put(-24,252){\Large\sf\black (c)}
	\put(-467,118){\Large\sf\white (d)}
	\put(-311,118){\Large\sf\white (e)}
	\put(-138,118){\Large\sf\white (f)}
	\put(-138,58){\Large\sf\white (g)}
	\caption{{\it Upper panels:} Statistical distribution of model field line apexes in the (a) pre-flare, (b) early-flare, and (c) late impulsive phase. Shown is the relative occurrence in \% of the number of considered field lines, within a certain time window (indicated by $t_{\rm start}$ and $t_{\rm end}$ in the top right corner of each panel). $\langle h\rangle$ denotes the median apex height ($\pm$ median absolute deviation) of the distribution (represented by a cross and a vertical line, respectively). A subset of the corresponding model field lines are shown in Figure~\ref{fig:fig5}b, \ref{fig:fig5}e, and \ref{fig:fig6}b, respectively. {\it Lower panels:} EUVI-A 195\,\AA\ images at (d) 07:25~UT and (e) 08:05~UT. The colored lines mark radially outward directed paths along which the intensity in the EUVI-A images is followed. Intensity stack plots along the blue dashed and yellow dashed-dotted line is shown as a function of time in panels (f) and (g), respectively. (A color version of this figure is available in the online journal.)\\}
	\label{fig:fig7}
\end{figure*}

\subsubsection{Height of the reconnection region}

Since we determined the coronal magnetic field structure of previously reconnected field, we can give a measure for how fast the reconnection site elevates in the model corona. More precisely, by tracing the model field lines rooted in flaring pixels, we can estimate a lower limit for the coronal altitude of the lower tip of the current sheet. In order to do so, we compute the apexes of the full set of field lines that are rooted in flare pixels, within the time intervals used for the discussion of the evolving coronal model field above (Sect.\ \ref{sss:mf_config}). Note that only a subset of these field lines is visualized in Figures~\ref{fig:fig5} and \ref{fig:fig6}.

For the pre- (early) flare magnetic field configuration we find a median apex height $\langle h\rangle$\,$\approx$\,$5\farcs0$ ($\approx$\,3.7\,Mm; Figure~\ref{fig:fig7}a). During phase {\sc I} of the flare (the first impulsive rise to C3-level), the median apex height of the magnetic field possibly involved in the magnetic reconnection process sometime before increases to $\langle h\rangle$\,$\approx$\,$9\farcs0$ ($\approx$\,6.6\,Mm; Figure~\ref{fig:fig7}b). During phase {\sc II} (until the flare peak M1-level) it further increases to $\langle h\rangle$\,$\approx$\,$19\farcs0$ ($\approx$\,14.0\,Mm; Figure~\ref{fig:fig7}c). From this, we estimate the elevation speed of the current sheet's lower tip as $\approx$\,3.6~km\,s$^{-1}$ during the impulsive phase (between $\approx$\,05:24~UT and $\approx$\,06:12~UT).

In time series of EUVI-A 195\,\AA\ images, we observe the successive formation of bright loops at successively higher apparent altitudes within several hours after the M1.4~flare (two time instances during the post-flare phase are shown in Figure~\ref{fig:fig7}d,e). Bright loops seen at this temperature are signatures of cooling of the initially hot flare loop plasma that was released from the reconnection region. Thus, also for the post-flare period, we are able to indirectly follow the elevation of the current sheet's lower tip indirectly. To this aim, we trace the intensity along several paths in the height range $1.0\le h\le1.1$~$R_{\rm Sun}$ (colored lines in Figure~\ref{fig:fig7}d,e). From the intensity values along the blue dashed path (Figure~\ref{fig:fig7}f), we find a two-stage evolution of the post-flare loop system, composed of a period of faster rise (with a velocity of $\approx$\,5.7~km\,s$^{-1}$) and a following slower one (rising with $\approx$\,2.2~km\,s$^{-1}$). The intensity along the yellow dashed-dotted path, suggests a mean velocity of $\approx$\,4.2~km\,s$^{-1}$ (Figure~\ref{fig:fig7}g). This also holds for slightly differing, radially outward directed paths but note that the coronal loops are seen in projection so that any deduced speed can only represent a rough estimate. Note that these velocity estimates are remarkably similar to that based on the (post-reconnection) NLFF model field line apexes during the impulsive phase (a few km\,s$^{-1}$).

The unambiguous identification of the post-reconnection loops in the EUVI-A 195\,\AA\ images is restricted to times after $\approx$\,06:48~UT (\ie, restricted to the post-flare phase). Before that time, the observed flaring emission is saturated and spatially not well resolved (see Figure~\ref{fig:fig7}f,g). We can assume that the highest apexes of flaring loops at that time are located at $h\gtrsim$\,$25\farcs0$ above the solar surface, but we are not able to trace the loop growth until that time in the EUVI-A 195\,\AA\ images. For comparison, NLFF model field lines rooted in flaring pixels allow us to estimate the elevation speed for the current sheet's lower tip in the pre- (early) flare to the late impulsive phase only (05:00--06:12~UT), since afterwards flare pixel information is sparse. However, taking the estimated median apex height estimated from the NLFF models at the end of the impulsive phase ($h\approx$\,$19\farcs0$ at 06:12~UT; see Figure~\ref{fig:fig7}c) and assuming that the post-reconnection field continues to elevate with a similar velocity as before ($\approx$\,3.6~km\,s$^{-1}$), the post-flare loop apexes should be located at a height of $\approx$\,$30\farcs0$ above the solar surface around 06:50~UT. This is in agreement with the flaring emission observed in the EUVI-A 195\,\AA\ images above the solar limb at that time.

\begin{figure*}
	\epsscale{1.0}
	\centering
	\includegraphics[width=2.0\columnwidth]{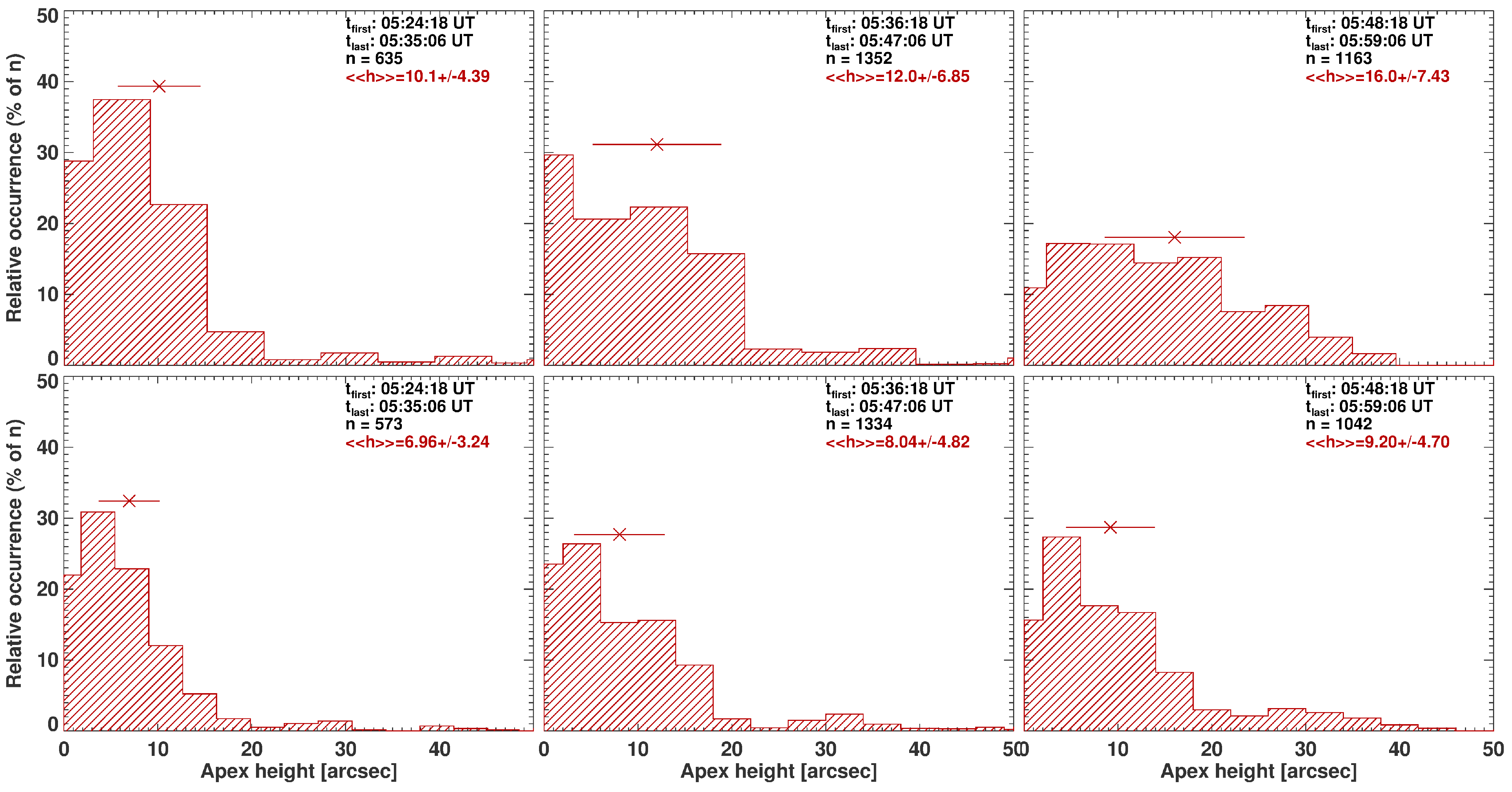}
	\put(-336,240){\Large\sf\black (a)}
	\put(-180,240){\Large\sf\black (b)}
	\put(-24,240){\Large\sf\black (c)}
	\put(-337,121){\Large\sf\black (d)}
	\put(-181,121){\Large\sf\black (e)}
	\put(-24,121){\Large\sf\black (f)}
	\caption{Statistical distribution of pre-flare ({\it upper panels}) and post-flare ({\it lower panels}) model field line apexes, calculated from the same flare pixel locations. Shown is the relative occurrence in \% of the number of considered field lines, calculated from the traced flare pixels within certain time windows during the impulsive phase of the flare (indicated by $t_{\rm start}$ and $t_{\rm end}$ in the top right corner of each panel). $\langle h\rangle$ denotes the median apex height ($\pm$ median absolute deviation) of the distribution (represented by a cross and a vertical line, respectively). (A color version of this figure is available in the online journal.)\\}
	\label{fig:fig8}
\end{figure*}

\begin{figure*}
	\epsscale{1.0}
	\centering
	\includegraphics[width=2.0\columnwidth]{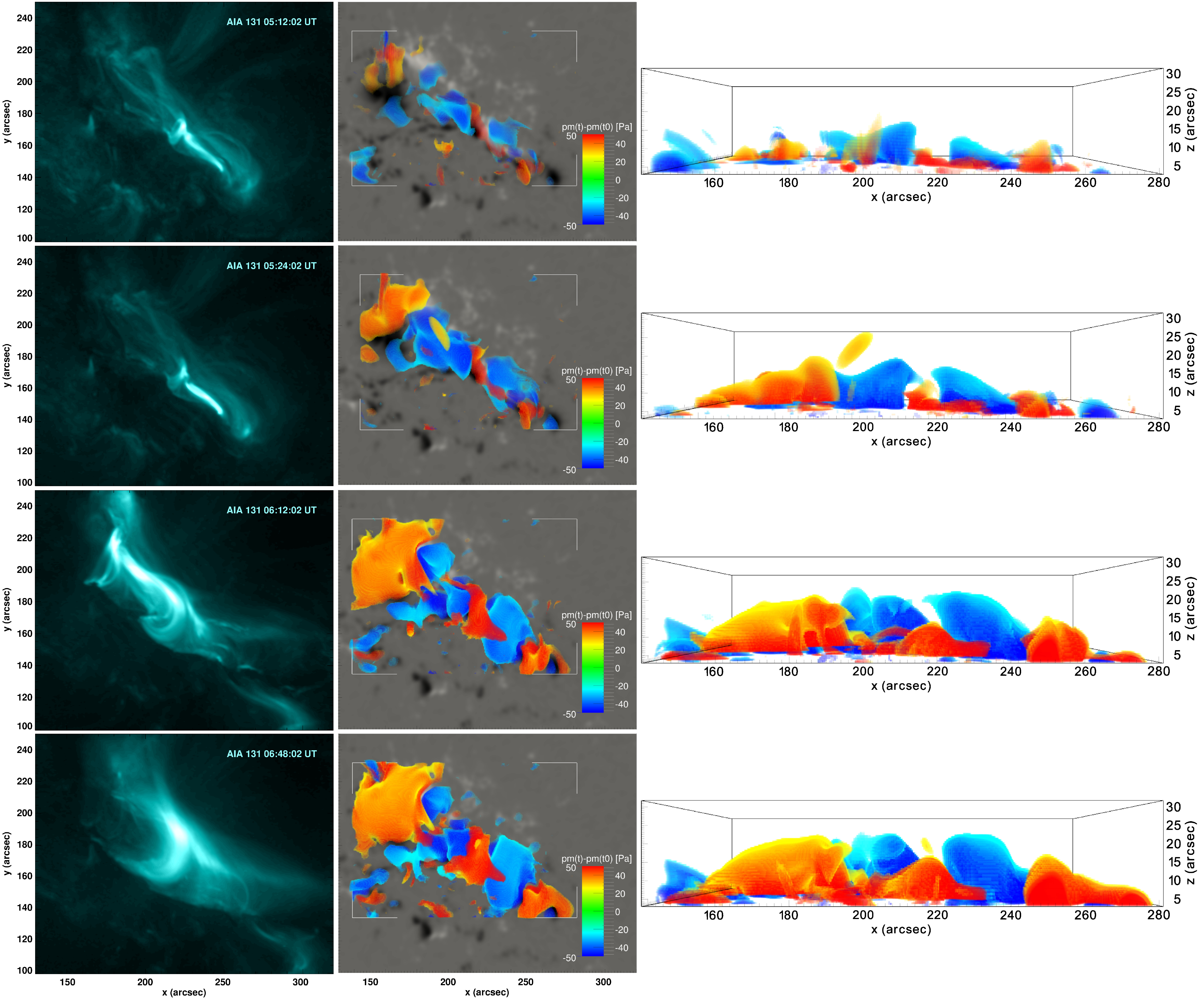}
	\put(-474,392){\Large\sf\white (a)}
	\put(-350,392){\Large\sf\white (b)}
	\put(-225,392){\Large\sf\black (c)}
	\put(-474,294){\Large\sf\white (d)}
	\put(-350,294){\Large\sf\white (e)}
	\put(-225,294){\Large\sf\black (f)}
	\put(-474,194){\Large\sf\white (g)}
	\put(-350,194){\Large\sf\white (h)}
	\put(-225,194){\Large\sf\black (i)}
	\put(-474,94){\Large\sf\white (j)}
	\put(-350,94){\Large\sf\white (k)}
	\put(-225,94){\Large\sf\black (m)}
	\caption{{\it Left panels:} AIA 131\,\AA\ emission, characteristic for the (a) pre-flare phase (05:12~UT), (d) early flare phase (05:24~UT), (g) late impulsive (06:12~UT) and (j) decay phase (07:00~UT). The FOV corresponds to the solid outline in Figure~\ref{fig:fig1}. {\it Middle panels:} Change of the magnetic pressure ${\rm d}p_m=p_m(t)-p_m(t_0)$ in the area associated to flaring activity (outlined by white corner elements) in the height range $3\farcs0\lesssim h \lesssim 31\farcs8$ (\ie, $2.2\lesssim h \lesssim 23.4$\,Mm). Red/blue corresponds to positive/negative values. Only values ${\rm d}p_m\ge25$~Pa are shown and values are scaled to $\pm$50~Pa. The gray-scale background resembles $\Bz$ on the NLFF lower boundary, co-temporal to the AIA 131\,\AA\ images and scaled to $\pm$2\,kG. {\it Right panels:} Same as middle panels but viewed along the positive $y$-direction, i.\,e., when looking towards solar north. Units are arcseconds from Sun center. (A color version of this figure is available in the online journal.)\\}
	\label{fig:fig9}
\end{figure*}

\subsubsection{Coronal implosion}\label{sss:implosion}

{During the impulsive phase of the M1.4~flare an expansion of the overall large-scale (potential) coronal loops can be observed in AIA EUV images, followed by the ejection of the CME-related material in south-western direction of the AR. That is spatially closely related to the low-lying highly twisted field connecting N2 and N3. As discussed above, these field structures close to the main PIL were apparently involved in reconnection during phase {\sc I} of the flare (between $\sim$05:19 and $\sim$06:00 UT; see middle column of Fig.~\ref{fig:fig5}). The tracing of the flare pixels within time intervals, partly covering this period, allows us to investigate the overall effect of the flare/CME process on the coronal magnetic field. As discussed in the introduction, any flare/CME associated release of previously stored magnetic energy should manifest itself in the form of an (partial) implosion of that portion of the coronal field that is directly involved in the reconfiguration process. 

In order to investigate the accompanied effect on the photospheric magnetic field, and consequently on the model lower boundary, we follow the same strategy as adopted by \cite{2015ApJ...804L..28S}. We first identify the PIL from a smoothed pre-flare photospheric magnetic field map (at 05:12~UT) and dilate it using a circular kernel. Similarly, we take the flare pixels tracked during phase {\sc I} of the flare and dilate that area with a large circular kernel. The overlap of the dilated PIL and flare pixel area -- the ``flare polarity inversion line'' (FPIL) -- is then used to investigate the flare-related changes on the lower boundary. We use five NLFF model fields prior to/after the nominal flare start/end in order to estimate the flare-related changes (covering $\approx$\,50~min before/after the flare, and where considering only locations hosting a magnetic field stronger than 100\,G). 

Compared to the mean pre-flare value, we find an average increase of $\approx$\,8\% of the mean horizontal field magnitude (compared to the pre-flare value $\langle\Bh\rangle=654\pm8.86$\,G). This finding supports the implosion conjecture, arguing for a less stressed (partially deflated) and less sheared magnetic field after the flare/CME. At the same time, we find a small ($\approx$\,2\%) decrease of the mean shear angle, compared to the corresponding pre-flare value ($\langle\theta\rangle=54.0\pm0.41^\circ$), which measures the angular difference between the actual field vector and its potential-field counterpart. Note that this value is to be interpreted in context with the ongoing shearing of the field along the main PIL during and after the analyzed period, as discussed in Sect.\ \ref{sss:preflare_evol}, which may compensate part of an eventually higher flare-induced decrease.

In a next step, we investigate the corresponding change of the overall field configuration also in the model volume above. To this aim, we trace field lines from a pre-flare and a post-flare NLFF model, starting from the same locations on the lower boundary. As start locations we chose all flare pixel locations tracked during phase {\sc I} of the flare (ranging from 05:24 to 06:00~UT; \ie, the time span within which the CME was initiated and the first impulsive energy release occurred). Here, we use the first/last of the series of pre-/post-flare NLFF model fields (at 04:24/07:36~UT). The statistics corresponding to the apexes of these pre- and post-flare model fields are shown in the upper and lower panels in Fig.~\ref{fig:fig8}, respectively. The flare pixels tracked during phase {\sc I} and used as start locations for field line calculation are, as before, separated into 12-min time intervals, in order to disentangle the effect on individual sets of field lines.

As can be seen from Fig.~\ref{fig:fig8}a--c, the median apex heights of the pre-flare model field lines range from $\approx$\,$10\farcs0$ at the beginning of phase {\sc I} to $\approx$\,$16\farcs0$ at the end of it. The field lines calculated from the same start locations in the post-flare model (Fig.~\ref{fig:fig8}d--f), on the other hand, show a clearly lower median apex height, being between $\approx$\,$3\farcs0$ ($\approx$\,2\,Mm) and $\approx$\,$7\farcs0$ ($\approx$\,5\,Mm) lower. This also implies that different portions of the coronal volume implode by a different amount. We note that the obtained values depend on the NLFF model used to calculate field lines from. The results presented here, however, is only slightly affected by using different pre-/post-flare models. The finding that the post-flare model field is more compact, reaching on average to lower coronal heights, supports that we indeed pictured the field previously involved in magnetic reconnection.
}

\subsubsection{Magnetic energy}

{In the previous section, we successfully identified the partial implosion of the coronal magnetic field, due to the sudden release of previously stored magnetic energy. Thus, we go a step further and analyze in more detail the coronal parts of the flaring AR which are associated to the loss and/or gain of magnetic energy in the course of the M1.4~flare under study. Therefore, we inspect the magnetic pressure, $p_m$, within the NLFF model volumes and address how the sub-volumes of energy loss/gain relate to the observed coronal EUV emission.} We calculate $p_m(t)=\mathf{B}(t)^2/(2\mu_0)$ for $t$=05:24~UT (early flare phase), $t$=06:12~UT (late impulsive phase), and $t$=07:00~UT (post-flare phase), and subtract it from the corresponding value of the pre-flare state (at $t_0$=05:00~UT). The resulting values at each point within the model volume indicate the change of the magnetic pressure, ${\rm d}p_m=p_m(t)-p_m(t_0)$. In Figure~\ref{fig:fig9}, we show the volume rendering of ${\rm d}p_m$ for $h\gtrsim$2.2\,Mm above the photosphere (corresponding to the height of the coronal base in our NLFF models).

Figure~\ref{fig:fig9}b indicates energy losses (\ie, ${\rm d}p_m$\,$<$\,0, indicated by blue color) co-spatial with bright coronal emission in the pre- (early) flare phase (Figure~\ref{fig:fig9}a). The pre-flare energy losses are found to be co-spatial with those during the flare, the latter covering larger and larger areas (compare to Figure~\ref{fig:fig9}e,h). Correspondingly, the locations at which magnetic energy is lost are found at (on average) larger heights in the atmosphere (compare Figure~\ref{fig:fig9}c to \ref{fig:fig9}f,i). The post-flare phase is characterized by energy losses comparable to those in the late impulsive phase (compare Figure~\ref{fig:fig9}k and h, as well as Figure~\ref{fig:fig9}m and i). {Noteworthy, the energy losses in the south-western part of the AR are found co-spatial with the coronal magnetic field structures for which an implosive character has been identified in Sect.~\ref{sss:implosion}.}

At the same time, magnetic energy is stored as well (${\rm d}p_m$\,$>$\,0, represented by red color). Already in the pre-flare phase, the system of low-lying highly sheared magnetic field connecting P2 and N3 (see Figure~\ref{fig:fig5}b) is associated to the storage of magnetic energy (Figure~\ref{fig:fig9}a). Note that the region in which energy is stored extends (on average) to lower heights than that at which magnetic energy is released. As later stages during the flare, energy storage is found more and more extended and successively higher in the model corona (compare Figure~\ref{fig:fig5}f, \ref{fig:fig5}i, and \ref{fig:fig5}m). Additional sources of energy storage behave similarly. One is located on top of the north-eastern part of the AR (partly covering N1/P1) and associated to the ongoing emergence of magnetic flux there. Another one is associated to the strongly sheared magnetic field between N3 and N4 in the south-west of the AR (compare Figure~\ref{fig:fig2}b).

\section{Discussion}

We analyzed the flare emission observed during a long-duration M1.4~flare {(SOL2011-08-02T06:19)}, hosted by NOAA~11261 on 2011 August~2, as well as its underlying coronal magnetic field configuration and evolution. We used coronal imagery and NLFF magnetic field modeling, respectively, for that purpose. We established a link between the observed flare emission and the coronal magnetic field in time and in space. The main findings can be summarized as follows.

Already in the pre- (early) flare phase, a distinct thermal \rhessi\ X-ray source was present, co-spatial with bright AIA 131\,\AA\ emission, and associated to highly twisted and/or sheared structures in the NLFF magnetic field models. Since no flare was recorded from this AR within hours before the analyzed M1.4~flare, we can exclude that the observed thermal source was a remnant of a preceding flaring activity. Thus, we interpret it as the signature of energy dissipation of electric currents stored in the sheared and/or twisted fields. The shear and/or twist was caused by the ongoing relative motion of the polarity patches within the AR. This relative motions persisted until days after the M1.4~flare and served as a constant source for the enhancement of shear and/or twist within the AR, thus played an important role for its flaring activity.

From the \goes\ SXR light curve, two phases were identified within the impulsive phase of the M1.4~flare. A first one characterized by the rise to C3 level (phase {\sc I}), followed by a second one when the emission rose to the peak (M1-) level (phase {\sc II}). Interestingly, the \rhessi\ X-ray sources associated to these two distinct (impulsive) phases were clearly separated in space and at different locations with respect to the bright pre-flare source. A corresponding spatial separation of the flaring loops in AIA EUV images was observed too. Flaring loops that connected the pre-flare structure to the south-west were observed in phase {\sc I}. Only during phase {\sc II}, flaring loops also connected to the north-east and were associated to the non-thermal \rhessi\ (footpoint) sources. This indicates that, only during the late stages of the impulsive phase, the plasma in the low solar atmosphere got heated due to the collision with flare-accelerated (non-thermal) electrons, propagating downwards from a coronal reconnection site.

{The model magnetic field associated to those locations in the low atmosphere which were previously involved in magnetic reconnection (identified by flaring pixels traced in AIA UV images), revealed a picture consistent with the standard model of eruptive flares. Starting from a highly sheared and/or twisted core region (close to the main polarity inversion line), the reconnection site evolves in coronal height, gradually involving the over-arching (more potential) field \citep[see review by][]{2002A&ARv..10..313P}.}

From the post-reconnection NLFF model field lines, rooted in flaring pixels, we deduced the growth of the post-reconnection system as a function of time, \ie, we estimated the elevation speed of the lower tip of the current sheet in the model corona. We found a velocity of $\lesssim$5\,km\,s$^{-1}$, during the impulsive phase of the M1.4~flare, supporting the findings of earlier studies of post-flare loops observed above the solar limb \citep[\eg,][]{2002SoPh..210..341G,2006SoPh..234..273V}, and similar to the growth rate deduced for post-flare giant arches \citep[][]{2015ApJ...801L...6W}. For the post-flare phase, we found a very similar estimate, based on the observed rise of the flare loop system in EUVI-A EUV images. This suggests an upward rise of the current sheet's lower tip at a comparable velocity during the impulsive and the post-flare phase. Importantly, the consistency of the NLFF model-based and observation-based estimate of the post-reconnection field's elevation, indicates that NLFF models may well be used for such a purpose, especially during the early stages of solar flares, when observations of a growing post-reconnection loop system suffer from saturated emission in the coronal images.

{The comparison of model field lines traced from a pre- and a post-flare NLFF model, originating from an identical set of start locations (defined by the flare pixels traced during phase {\sc I} of the flare), allowed us to reconstruct the partial implosion of the coronal volume above the flare site, supporting the conjecture put forward by \cite{2000ApJ...531L..75H}. We found a more compact post-flare magnetic field, compared to the pre-flare state (with a few Mm lower apex height, on average). The expected corresponding increase in the lower boundary's (photospheric) average horizontal magnetic field magnitude was found as $\approx$\,8\% ($\approx$\,50~G). This is in line with earlier model-based findings on that topic as of, \eg\ \cite{2015ApJ...804L..28S}, who found an increase of $\gtrsim$\,15\% for two major (\goes\ class $>$\,X2) eruptive flares. Assuming the flare-induced changes to scale with the size of the event, the values found in our study seems reasonable, given the size of the event analyzed (\goes\ class M1).}

{In line with the above discussed changes to the quasi-static model magnetic field in and around NOAA~11261, we found a mix of sub-volumes within which magnetic energy was locally stored or released.} In the course of the impulsive phase, the losses of magnetic energy appear to be located at successively larger heights, whereas they appear to remain at comparable heights during the decay phase. They spatially correlate with bright flaring emission in AIA EUV images {and that portion of the coronal model volume which had been shown to partially implode}, supporting that the bright coronal emission points to the places in the coronal volume that are associated to the conversion of magnetic energy into kinetic energy and/or heat. In accordance to earlier studies \citep[\eg,][]{2012ApJ...748...77S,2014JGRA..119.3286H}, we found the energy losses at low heights in the corona, predominantly below $\approx$\,15\,Mm.

We found that the volume above ongoing photospheric flux emergence and increasingly sheared and/or twisted magnetic fields (the latter established by the relative motion of magnetic structures) were dominated by the successive storage of magnetic energy. This underlines that the consequences of the relative motion of photospheric magnetic structures within the AR, in combination with flux emergence and cancellation, determine the fate of solar ARs from their early stages on. Moreover, we found that the energy storage continued to large heights even after the flare, in contrast to the energy losses, apparently remaining at heights similar to those of the impulsive phase.

\acknowledgments
\small
{We thank the anonymous referee for careful consideration and helpful comments.} J.\,K.\,T.~acknowledges financial support by Austrian Science Fund (FWF): P25383-N27. A.\,V.~and Y.\,S.~acknowledge financial support by Austrian Science Fund (FWF): P27292-N20. Y.\,S.\ also acknowledges the Thousand Young Talents Plan, a sub-program of the ``Recruitment Program of Global Experts'' (1000 Talent Plan) and 11233008 from National Natural Science Foundation of China (NNSFC). SDO is a mission for NASA’s Living With a Star (LWS) Program. \sdo\ data are courtesy of the NASA/\sdo\ and HMI science team.\\

\bibliographystyle{apj}
\bibliography{bibliography}

\end{document}